\documentclass{llncs}

\usepackage{latexsym}
\usepackage{listings}
\usepackage{algorithmicx,algpseudocode}
\usepackage{tikz}
\usepackage{multirow}
\usepackage{color}
\usepackage{array}
\usepackage{url}
\usepackage{multirow}
\usepackage{amssymb}
\usepackage{comment}
\usepackage{rotating}
\usepackage{footnote}
\usepackage{censor}
\makesavenoteenv{tabular}

\newcommand{\COMPR}{{\mathcal C}}
\usetikzlibrary{shapes,positioning}
\def\layersep{1.6cm}

\begin{document}

\mainmatter 

\title{Neural Networks as Artificial Specifications \vspace{-5mm}}
\author{
\vspace{-3mm}
I.S.W.B. Prasetya\orcidID{0000-0002-3421-4635} \and Minh An Tran}
\institute{
  Utrecht University,
  the Netherlands, \email{s.w.b.prasetya@uu.nl}
}

%\toctitle{Lecture Notes in Computer Science}
%\tocauthor{Authors' Instructions}
\maketitle

\vspace{-7mm}
\begin{abstract}
  In theory, a neural network can be trained to act as an artificial specification for a program
  by showing it samples of the programs executions. In practice, the training turns out to be
  very hard. Programs often operate on discrete domains for which patterns are difficult to discern.
  Earlier experiments reported too much false positives. This paper revisits an experiment by Vanmali et al. 
  by investigating several aspects that were uninvestigated in the original work:
  the impact of using different learning modes, aggressiveness levels,
  and abstraction functions. The results are quite promising.
\end{abstract}

\vspace{-8mm}
  
\

\noindent
{\scriptsize NOTICE: This is a pre-print of the same paper published in the proceedings of
the 30th International Conference on Testing Software and Systems (ICTSS) 2018,
LNCS. Publisher: Springer-Nature. DOI: \url{https://doi.org/10.1007/978-3-319-99927-2_11}.
Additional results have also been added in the Appendix which were not included
in the published version due to the space limitation.
}  

\

%=====================================================================
\vspace{-5mm}
\section{Introduction}
%=====================================================================

\vspace{-4mm}
Nowadays, many systems make use of external services or components to do some of their tasks,
allowing services to be shared, hence reducing cost. However, we
also need to take into account that third parties services may be updated on the fly as
our system is running in production. If such an update introduces an error, this may affect the correctness
of our system as well. One way to guard against this is by doing run time verification \cite{cao2010automated}:
at the runtime the outputs of these services are checked against their formal specifications.
Unfortunately, in practice it is hard to persuade developers to write formal specifications. 

A more pragmatic idea is to use 'artificial specifications' generated by a computer. 
Another use case is automated testing. Tools like
QuickCheck, Evosuite, and T3 \cite{QuickCheck,fraser2011evosuite,T3i} 
are able to generate test inputs, but if no specification is given, only common correctness conditions such as 
absence of crashes can be checked. Using artificial specifications would extend their range.

Although we cannot expect a computer to be able to on its own specify the intent of a program, 
it can still try to guess this intent. 
One way to do this is by observing some training executions to predict
general properties of the program, e.g. in the form of 'invariants' (state properties)
\cite{ernst2007daikon}, finite state machine \cite{mariani2008automated}, 
or algebraic properties \cite{ABBA}.
These approaches cannot however capture the full functionality of a program, e.g.
\cite{ernst2007daikon} can only infer predefined families of predicates,
many are simple predicates such as such as $o{\not=}{\sf null}$ and $x{+}y{\geq}0$.
With respect to these approaches, neural networks offer
an interesting alternative, since they can be trained to simulate a function
\cite{kriesel2007brief}. 

The trade off of using artificial specifications is the additional overhead in debugging. 
When a production-time execution violates such a specification, 
the failure may be either caused by an error triggered by the execution, or by an error in the training executions that were
reflected in the predictions, or due to inaccuracy of the predictions. The first two cases 
expose errors (though the second case would take more effort to debug). However, the failure in the 
last case is a false alarm (false positive). Since we do not know upfront if a violation
is a real error or a false positive, we will need to investigate it (debugging), which is
quite labour intensive. If it turns out to be a false positive, the effort is wasted. 
Despite the potential, studies on the use of neural networks
as artificial specifications are few:
%\cite{vanmali2002,aggarwal2004neural,agarwal2004comparative,mao2006neural,lu2007oracle}.
\cite{vanmali2002,aggarwal2004neural,mao2006neural,lu2007oracle}.
They either reported unacceptably high rate of false positives, or do not address the issue.

In this paper we revisit an experiment by Vanmali et al. \cite{vanmali2002} that 
revealed $\approx 16\%$ rate of false positives ---a rate of above 5\% is likely to render any 
approach unusable in practice.
The challenge lies in the discrete nature of the program used as the experiment subject, making it 
very hard to train a neural network.
This paper explores several aspects that were left uninvestigated in the original work,
namely the influence of different learning modes, aggresiveness levels, and abstraction.
The results are quite promising.
% ---these should be considered as preliminary since
%we are yet to verify them on more subjects; nevertheless, 
%we believe that there are useful insights that can already be learned.

%=====================================================================
\vspace{-4mm}
\section{Neural Network as an Artificial Specification} \label{sec.NN}
%=====================================================================

\vspace{-4mm}
Consider a program $P$ that behave as a function $I {\rightarrow} O$. 
An artificial {\em specification} $\phi$ is a predicate $I {\times} O
{\rightarrow} {\sf bool}$; $\phi(x,P(x)){=} {\sf T}$ means that $P$'s output is judged as correct, and else incorrect.
With respect to the intended specification $\mathcal G$, $\phi$'s judgment is
a {\em true  positive} is when both $\phi$ and $\mathcal G$ judge a {\sf T},
a {\em true  negative} is when they agree on the judgement {\sf F},
a {\em false  positive} is when $\phi$ judges $\sf F$ and $\mathcal G$ judges $\sf T$,
and a {\em false  negative} is when $\phi$ judges $\sf T$ and $\mathcal G$ judges $\sf F$.

%Let's now consider using an artificial neural network (NN) as an artificial specification for $P$.
An {\em neural network} (NN) is a network of '{\em neurons}' \cite{kriesel2007brief} that behaves as a function $\mathbb{R}^M {\rightarrow}
\mathbb{R}^N$.
We will restrict ourselves to {\em feed forward NNs} (FNNs) where the
neurons are organized in linearly ordered layers \cite{kriesel2007brief}; an example is below: 
%
%
%\begin{figure}
  \vspace{-2mm}
  \begin{center}
    \begin{tikzpicture}[shorten >=1pt,->,draw=black!90, node distance=\layersep]
    \tikzstyle{every pin edge}=[<-,shorten <=1pt]
    \tikzstyle{neuron}=[circle,draw,fill=black!25,minimum size=5pt,inner sep=0pt]
    \tikzstyle{input neuron}=[neuron, fill=green!50];
    \tikzstyle{output neuron}=[neuron, fill=red!50];
    \tikzstyle{hidden neuron}=[neuron, fill=blue!50];

    % Draw the input layer nodes
    \foreach \name / \y in {0,...,2}
        \node[input neuron, pin=left:In $I_\y$] (I\name) at (0, 1 - 0.4*\y) {};

    % Draw the hidden layer nodes
    \foreach \name / \y in {0,...,3}
        %\path[yshift=0.4cm]
        \node[hidden neuron] (H\name) at (\layersep, 0.4*\y) {};

    % Draw the output layer node
    % \node[output neuron,pin={[pin edge={->}]right:Output}, right of=H-2] (O) {};
    \node[output neuron,pin={[pin edge={->}]right:$O_0$}] (Out0) at (2*\layersep,1) {};
    \node[output neuron,pin={[pin edge={->}]right:$O_1$}] (Out1) at (2*\layersep,0.2) {};

    % Connect every node in the input layer with every node in the
    % hidden layer.
    \foreach \source in {0,...,2}
        \foreach \dest in {0,...,3}
            \path (I\source) edge (H\dest);

    % Connect every node in the hidden layer with the output layer
    \foreach \source in {0,...,3}
             { \path (H\source) edge (Out0);
               \path (H\source) edge (Out1); }

\end{tikzpicture}

  \end{center}
%  \vspace{-5mm}
%  \caption{\em An example of an FNN consisting of three layers. 
%  } \label{fig.example.FNN}
%  \vspace{-5mm}
%\end{figure}
%
\vspace{-3mm}
The first layer is
called the {\em input layer}, consisting of $M$ neurons connected to the inputs. The last layer is the
{\em output layer}, consisting of $N$ neurons that produce the outputs.
The layers in between are called {\em hidden layers}. 
An input neuron simply passes on its input, else it
has $k$ inputs and an additional input called 'bias' whose value is always 1 \cite{kriesel2007brief}.
%\footnote{Together with its weight, a bias
%  input is used to represent a threshold that the rest of
%  the neuron's inputs must collectively overcome.}.
%
Each input connector has a weight $w_i$.  The neuron's output is the weighted sum of its inputs, 
followed by applying a so-called {\em activation function}:
$ out = f\;({\large\rm\Sigma}_{0{\leq}i{\leq}k} \ w_i . x_i)$.
A commonly used $f$ is the logistic function, which we also use in our experiments.

Any continuous numeric function $\mathbb{R}^M {\rightarrow}
\mathbb{R}^N$, restricted within any closed subset of 
$\mathbb{R}^M$, can be simulated with arbitrary accuracy by an FNN \cite{goodfellow2016deep},
which implies that an FNN can indeed act as an artificial specification for $P$, if $P$ is
injectable into such a numeric function. That is, there exists a continuous numeric function 
$F{:}\mathbb{R}^M {\rightarrow} \mathbb{R}^N$ and injections $\pi_I {:} I {\rightarrow } \mathbb{R}^N$
and $\pi_O {:} O {\rightarrow } \mathbb{R}^N$
such that $F$ encodes $P$: for all $x{\in}I$, $P(x) {=} \pi_O^{-1}(F(\pi_I(x)))$.
%with enough layers and neurons \cite{goodfellow2016deep}. 
%Functions over discrete domains can be approximated by mapping the domains to $\mathbb{R}^k$.  
However, finding a right FNN is hard. A common technique to find one
is by training an FNN using a set of sample inputs and outputs, e.g. using
the back propagation \cite{kriesel2007brief}  algorithm. 
It might be easier to train the NN to simulate $\alpha \circ P$ instead, where $\alpha$ is some chosen abstraction on $P$'s output values.
The trade off is that we get a weaker specification.

Since an NN does not literally produce a {\sf bool}, 
we couple its output vector $\bar{z}' {=} {\rm NN}(\pi_I(\bar{x}))$ to a so-called {\em comparator}
$\COMPR:\mathbb{R}^N {\rightarrow} \mathbb{R}^N {\rightarrow} {\sf bool}$ to
calculate the judgement by comparing $\bar{z}'$ with the observed output $\bar{z}{=}\pi_O(\alpha(P(\bar{x})))$. 
Basically, if their values are 'far' from each other, then the judgement is
$\sf F$, and else $\sf T$. By adjusting what 'far' means we can tune the specification's aggressiveness
without having to tamper with the NN's internals. In our experiments (below), 
the identity function $id {=} (\lambda x.\; x)$ will be used as 
the injector $\pi_I$ and $\pi_O$.  
Because $id$ simply passes on its input, it will be omitted from the formulas.

%=====================================================================
\vspace{-5mm}
\section{Experiments} \label{sec.experiment}
%=====================================================================

\vspace{-4mm}
Figure \ref{fig.subject} shows a credit approval program from the financial domain that was used as the 
experiment subject by Vanmali et al \cite{vanmali2002}. 
%
%, where FNN produced a false positive rate of 16\%. We want to investigate if there is a way to improve the FNN.
%
The program takes 8 input parameters describing a customer. 
The output is a pair $(b,y)$ where $b$ is a boolean indicating whether
the credit request is approved, and if so $y$ specifies the maximum allowed credit. 
We will ignore $b$ since \cite{vanmali2002} already shows that 
an FNN can accurately predict its value.
%, and will instead focus on constructing an artificial specification for $y$.
%
Despite its size, the subject is quite challenging for an NN to simulate because it operates on
a discrete domain (the numeric values are all integers). The whole input domain has 224000 possible values.
We will use an FNN with 8 inputs (representing $\sf approve$'s inputs) and a hidden layer with 24 neurons
(adding more layers and neurons does not really improve the FNN's accuracy).

%The range of values that each parameter can take is shown below.
%
%\begin{center}
%{\footnotesize
%  \begin{tabular}{|lc|lc|} \hline
%   {\em parameter} & {\em range}  \\ \hline
%   \lstinline$Citizenship$ & 0:US, 1:other \\ \hline
%   \lstinline$State$       & 0:Florida,  1:other \\ \hline
%   \lstinline$Age$         & integer 1..100 \\ \hline
%   \lstinline$Sex$         & 0:female, 1:Male \\ \hline
%   \lstinline$Region$      & a region in US: integer 0..6 \\ \hline
%   \lstinline$Income$      & income class: integer 0..3 \\ \hline
%   \lstinline$Dependents$  & \# dependent family members: integer 0..4 \\ \hline
%   \lstinline$Marital$     & 0:single, 1:married \\ \hline
%   \lstinline$Approved$    & 0:credit rejected, 1:approved &
%   \lstinline$Amount$      & credit limit: integer $0..18000$ \\ \hline 
%\end{tabular}
%}
%\end{center}

\begin{figure}[t]
\begin{lstlisting}[numbers=left, numberstyle=\tiny, numbersep=5pt, xleftmargin=15pt]
approve(Citizenship,State,Region,Sex,Age, Marital,Dependents,Income) {
  if(Region==5 || Region==6) Amount=0 ; 
  else if(Age<18) Amount=0 ;
  else {
  if(Citizenship==0) {
     Amount = 5000+1000*Income ;
     if(State==0)
        if(Region==3 || Region==4) Amount = Amount*2 ;
        else  Amount = (int)(Amount*1.50) ;
     else Amount = (int)(Amount*1.10) ;
     if(Marital==0)
        if(Dependents>0) Amount = Amount+200*Dependents ;
        else Amount = Amount+500;
     else Amount = Amount+1000 ;
     if(Sex==0) Amount = Amount+500 ;
     else Amount = Amount+1000;
  }
  else {
     Amount = 1000 + 800 * Income;
     if(Marital==0)
        if(Dependents>2) Amount = Amount+100*Dependents ;
        else Amount = Amount+100 ;
     else Amount = Amount+300 ;
     if(Sex==0) Amount = Amount+100 ;
     else Amount = Amount+200 ;
  }
  if(Amount==0) Approved=F else Approved=T;
  return (Approved,Amount); }
\end{lstlisting}
\vspace{-4mm}
\caption{\em The experiment subject: a credit approval program from \cite{vanmali2002}.} \label{fig.subject}
\vspace{-5mm}
\end{figure}

Five variations of the FNN will be used, as listed below, along with the used comparator $\COMPR$.
 $\COMPR$ is parameterized with aggressiveness level $A$  (integer 0 (least aggressive) ... 5) that determines 
$\COMPR$'s policy to deal with non clear-cut cases.

\begin{enumerate}
    \item The FNN $\sf direct$ has one output, which is trained to simulate $y$.  Its comparator $\COMPR_A$ uses Euclidian distance,
    with sensitivity linearly scaled by $A$:
       $\COMPR_A(y,y') \ = \ |y - y'| < \epsilon_{max} - 0.01 A$, with $\epsilon_{max} {=} 0.09$.  
       
    \item The FNN ${\sf uni}_N$ has $N$ outputs, trained to simulate $\alpha_N \circ \; {\sf approve}$. The abstraction
    $\alpha_N$ maps {\sf approve}'s $y$ output to a vector $\bar{z} : [0.0..1.0]^N$ representing one of $N$ uniform sized intervals 
    in $y$'s range [0..18000], such that the $k$-th interval is represented by a vector of 0's except a single 1 at the $k$-th position. 
    If $\bar{v} : [0.0..1.0]^N$, let ${\sf winner}(\bar{v})$ be the index of the greatest element in $\bar{v}$. 
    
    The comparator is more complicated. An obvious case is when $\bar{z'} = {\rm NN}(\bar{x})$ and 
    $\bar{z} = \alpha_{10}({\sf approve}(\bar{x}))$      
    report the same winner. If the NN's winner is confident of itself,
    $\sf approve$'s output is judged as correct.
    When they produce different winners and the NN's winner is confident of itself, we judge
    $\sf approve$ to be incorrect. Other cases are non-clear-cut and judged depending on the aggressiveness level.
    The full definition of $\COMPR_A$ is shown below. The original work Vanmali et al. \cite{vanmali2002} only uses
    $A=3$ aggressiveness level.
    \begin{center}\scriptsize
    \vspace{-2mm}
    \begin{algorithmic}
      \Function{$\COMPR_A$}{$\bar{z}, \bar{z}'$}
      %\State {$\triangleright$ $g$ specifies the aggresiveness level}
      \State $k,j \gets {\sf winner}(\bar{z}),{\sf winner}(\bar{z}')$ \ \ ; \ $agree \gets k=j$
      \State ${\bf if} \hspace{10mm} agree \; \wedge \; |agree {-} \bar{z}'_j| < th_{low} \hspace{2mm} {\bf then} 
                                          \; \mbox{(obvious match)} \; {\sf T}$ 
      \State $\mbox{\bf else if} \; \neg agree \; \wedge \; |agree {-} \bar{z}'_j| > th_{high} \ {\bf then} 
                                          \; \mbox{(obvious mismatch)}\; {\sf F}$ 
      \State ${\bf else} \; \mbox{(non-clear-cut cases)} \; {\bf case} \; A \; {\bf of}$ 
      \State $\begin{array}{lll}
          0: & \mbox{(least aggressive: always accept)} \
             {\sf T}
             \\   %TTTT (match-medium, match-far, nonmatch-medium, nonmatch-close)
          1: & \mbox{(reject when the NN contradicts agreement)} \
              \neg (agree \wedge |{\sf T} {-} \bar{z}'_j| > th_{high})
             \\   %TFTT      
          % 2: alarm \gets (winnerMatch \wedge \delta{=}far) \vee (\neg winnerMatch \wedge \delta{\geq}medium) \\   %TFFT
          % changing 2 this to:
          2: & \mbox{(always accept on agreement)} \
             agree
             \\   %TFFT
          3: & \mbox{(Vanmali et al. \cite{vanmali2002}: accept on conflicting results)} \
             \neg agree \vee |{\sf T} - \bar{z}'_j| > th_{high} 
             %& {\bf return}\; winnerMatch \wedge \delta{=}medium 
             \\   %FTTT
          % 4: alarm \gets (winnerMatch \wedge \delta{\geq}medium) \vee (\neg winnerMatch \wedge \delta{\geq}medium) \\   %FFFT
          % 4 can be simplified to:
          4: & \mbox{(only accept if NN's winner supports $\bar{z}$)} \
             |agree - \bar{z}'_j| < th_{low} 
             \\   %FFFT
          5: & \mbox{(most aggressive: never accept)}\
             {\sf F} 
             \\   %FFFF
        \end{array}$
      \EndFunction
    \end{algorithmic}
    \end{center}
    \vspace{-2mm}
    The thresholds $th_{low}$ and $th_{high}$ are set to $0.2/0.8$. 
    
    \item The FNN ${\sf unimin}_N$ is a less presumptuous variant of $\sf uni$, with $th_{low}/th_{high}$
    set to $0.1/0.9$. This will cause more cases to be regarded as non-clear-cut.

    \item The FNN ${\sf lower}_N$ is like ${\sf uni}_N$, but trained to simulate $\alpha_N \; \circ \; low \; \circ {\sf approve}$.
    $low$ is used to 'stretch' $\alpha_N$ to divide $y$ into finer intervals in the lower region of $y$'s range, 
    e.g. if we believe the region to be more error prone, and growing coarser towards 
    the other end. 
    We use the log function to do this: $K * log(1 + y/a)$ with $K{=}8000$ and $a{=}100$ controlling the steepness.

    \item The FNN ${\sf center}_N$ is like ${\sf uni}_N$, but trained to simulate $\alpha_N \; \circ \; ctr \; \circ \; {\sf approve}$.
    $ctr$ is used to 'stretch' $\alpha_N$ to divide $y$ into finer intervals in the center region of $y$'s range. 
    We use logistic function 
    $ctr(y) {=} M / (1 + e^{-a(y - 0.5 M)})$ where $M{=}18000$ ($y$'s maximum) and
    $a{=}0.0006$ determines the function's steepness.

\end{enumerate}

{\bf Training}. We randomly generate 500 distinct inputs (from
 the space of 224000 values) and collect the corresponding ${\sf approve}$'s outputs.
 This set of 500 pairs (input,output) forms the training data. For every type of
 FNN above and every aggressiveness level an FNN is trained. 
 $N$ controls the granularity of the used abstraction, so we also try various $N$ (10..60).
 For each FNN, the connections' weight is randomly initialized  
 in $[-0.5..0.5]$. The training is done in a series of epochs using the back
 propagation algorithm \cite{kriesel2007brief}. We tried both 
 the incremental learning mode \cite{kriesel2007brief,FANNCsharp},
 where the FNN's error is propagated back after each training input,
 and batch learning modes, where only the average error is propagated back, after the whole batch of training inputs (500 of them).
 Incremental learning is thus more sensitive to the influence of individual inputs. 
 %Batch learning propagates the average of the FNN's errors over the whole batch, so in theory it is more steady.
 
{\bf Evaluation}. To evaluate the FNNs' ability to detect errors, we run them
on 21 erroneous variations (mutants) of the subject as in \cite{vanmali2002}
---they are listed in the Appendix.
%---due to limited space they are not shown here.
For each mutant, 500 distinct random inputs are generated, whose outputs are 'error exposing'
(distinguishable from the corresponding outputs of the correct subject). As an artificial specification, an
FNN should ideally reject all these error exposing outputs. Each rejection is a true positive.
We also generate 500 distinct random inputs and feed it to the (unmutated) subject. The FNN should
accepts the corresponding outputs ---each rejection is a false positive.

%=====================================================================
%\vspace{-4mm}
%\section{Analysis} \label{sec.analysis}
%=====================================================================

\begin{figure*}
  \vspace{-7mm}    
  \begin{center}
  %\vspace{-22mm}
  %\begin{tabular}{l}
  %\includegraphics[trim={30mm 22cm 5cm 9mm},clip,scale=0.35]{plot1_TP.png} 
  \includegraphics[trim={7mm 0mm 0mm 0mm},clip,scale=0.35]{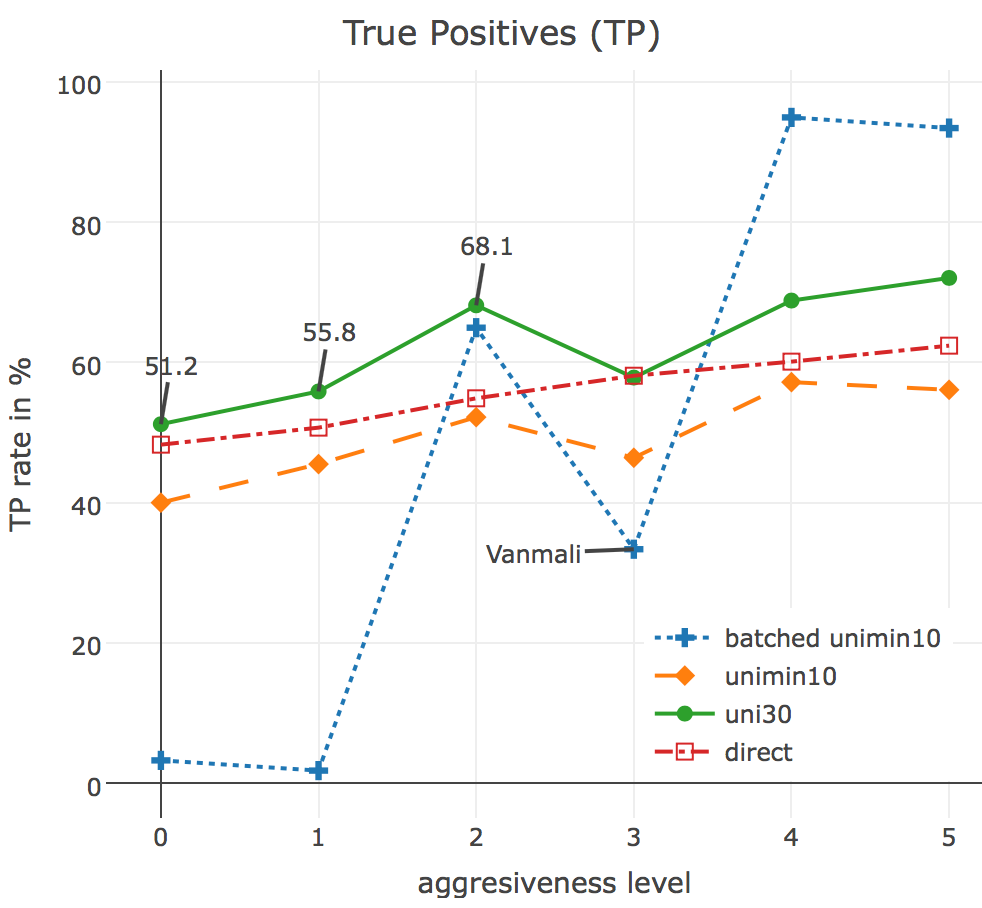} 
  \hspace{3mm}
  \includegraphics[trim={7mm 0mm 0mm 0mm},clip,scale=0.35]{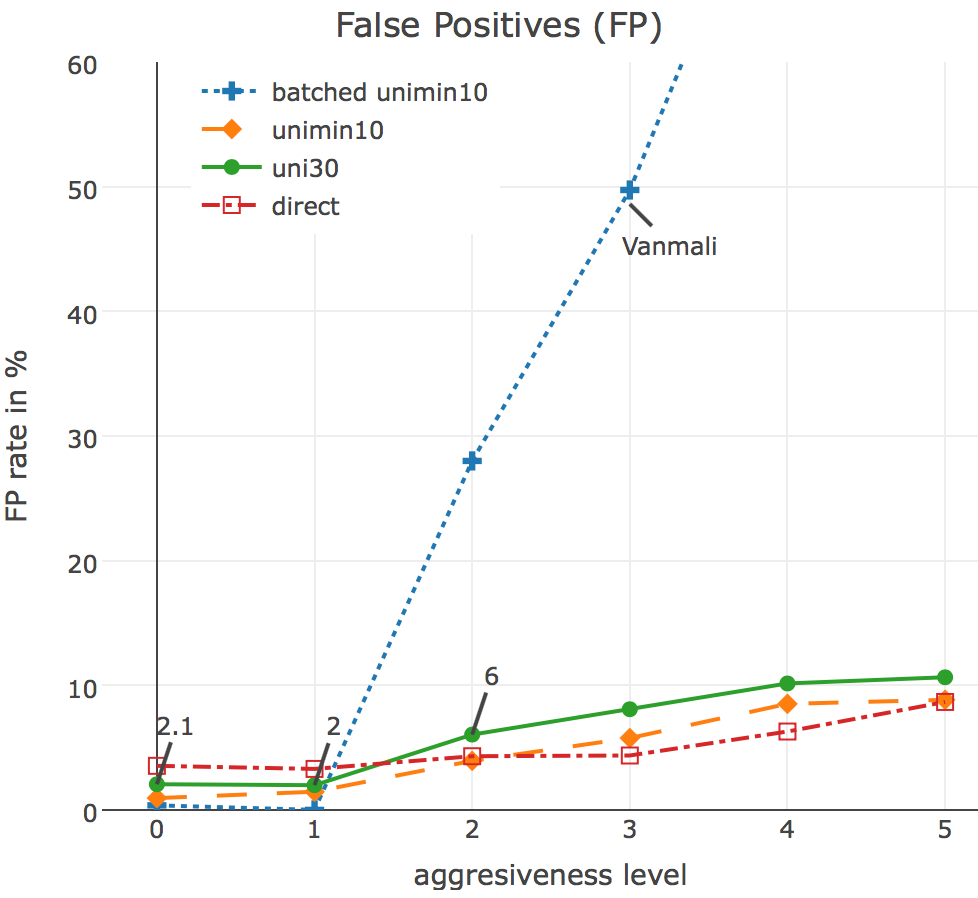} 
  %\end{tabular}
  \end{center}
  \vspace{-6mm}
  \caption{\em The true positive and false positive rates (in \%) of different FNNs.}
  \label{fig.TPFP.plot1}
\end{figure*}

\vspace{-6.5mm}
Figure \ref{fig.TPFP.plot1} shows some of the results. Except for $\sf direct$,
the training was done in 1500 epochs with learning rate 0.5.
We can see that using abstraction improves the FNN's performance: compare $\sf direct$ with ${\sf uni}_{30}$. 
The latter obtains a true positive rate 68\% on aggressiveness 2, implying
that out of two erroneous executions, ${\sf uni}_{30}$ is likely to detect at least one,
while when the aggressiveness level is set low, its rate of false positives is only around 2\%. Abstraction also makes training easier:
after 1500 epochs ${\sf uni}_{30}$ produces a mean square error (MSE) of $\approx 0.0001$,
whereas the shown results for $\sf direct$ is obtained after 10000 epochs (incrementally) with 0.1 learning rate, 
yielding an MSE $\approx 0.0004$.

The experiment in \cite{vanmali2002} uses ${\sf unimin}_{10}$. We believe
\cite{vanmali2002} used batch learning because the reported MSE after 1500 epochs matches, namely $\approx 0.05$.
However, as can be seen in Figure \ref{fig.TPFP.plot1}, this leads to poor performance (${\sf batched\;unimin}_{10}$).
Incremental learning yields a much more accurate FNN ($\approx 0.0001$ MSE), hence also
better performance (${\sf unimin}_{10}$).
%Our evaluation setup is not exactly the same as \cite{vanmali2002}. We use equal number
%of correct and erroneous executions (500 + 500) per mutant, whereas \cite{vanmali2002}
%uses in total 1000 executions, but the distribution between correct and erroneous executions per mutant
%seems to be arbitrary. 
The performance of the FNN in \cite{vanmali2002} under our setup is
indicated by the $\sf vanmali$-markers in Figure \ref{fig.TPFP.plot1}. 

\begin{figure*}
  \vspace{-5mm}
  \begin{center}
  %\begin{tabular}{lr}
  %\includegraphics[trim={30mm 22cm 5cm 0},clip,scale=0.35]{plot2_TP.pdf} & % \hspace{-5mm}
  %\includegraphics[trim={30mm 22cm 5cm 0},clip,scale=0.35]{plot2_FP.pdf}
  \includegraphics[trim={7mm 0mm 0mm 0mm},clip,scale=0.34]{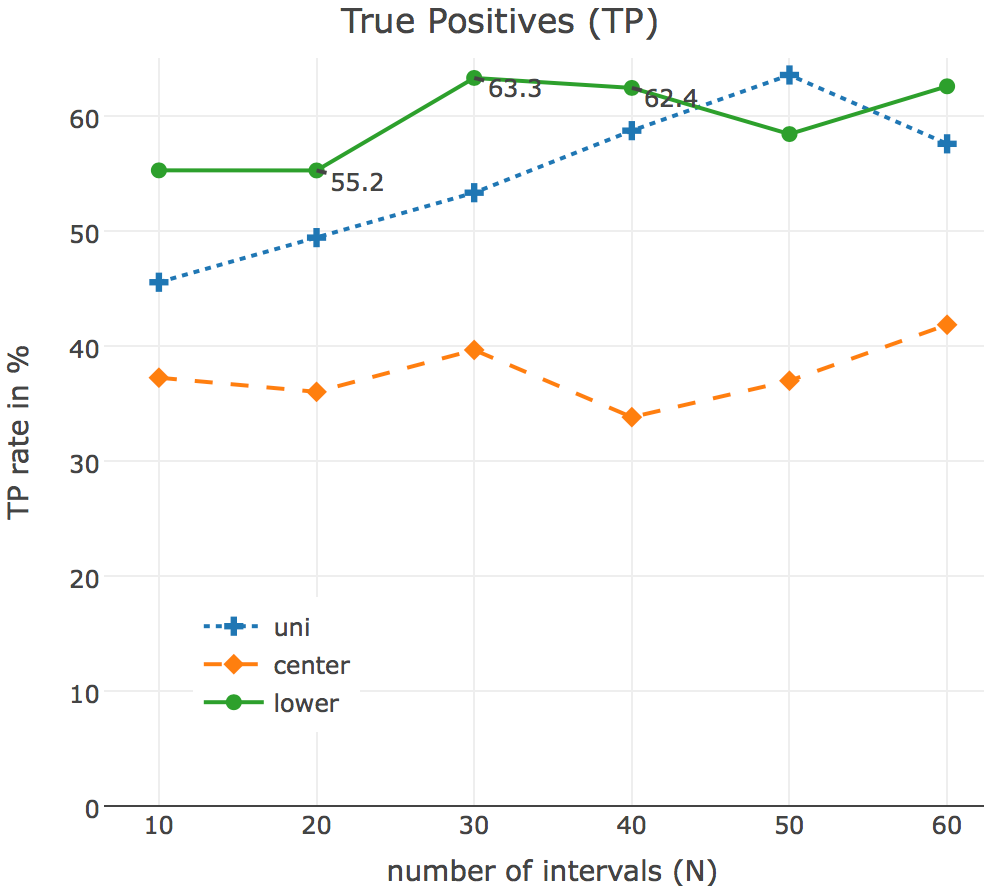} 
  \hspace{3mm}
  \includegraphics[trim={7mm 0mm 0mm 0mm},clip,scale=0.34]{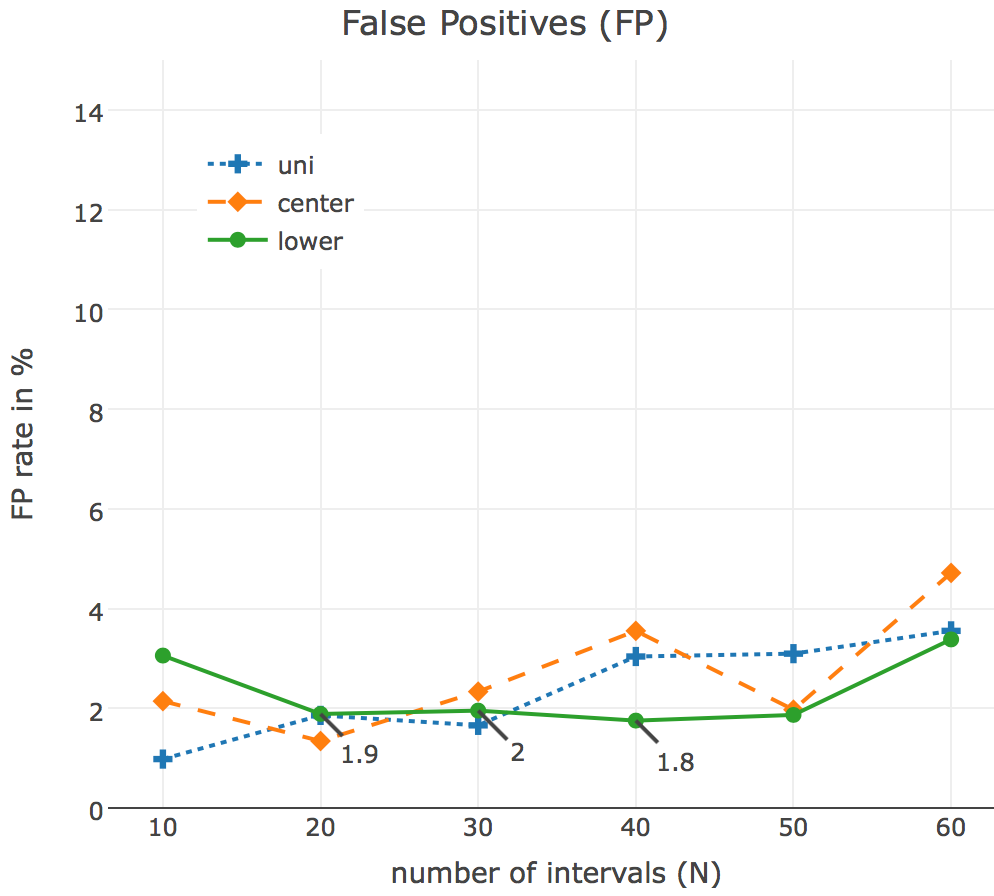} 
  %\end{tabular}
  \end{center}
  \vspace{-6mm}
  \caption{\em The effect of different abstractions and the abstraction granularity ($N$).
   $\sf uni$ shows the TP and FP rates of the ${\sf uni}_N$ configuration with its aggressiveness 
   level set to 0 ---recall that this FNN uses the function $\alpha_N$ as abstraction.
   $\sf center$ and $\sf lower$ show the TP and FP rates of the same FNN, but they use respectively $\alpha_N \; \circ \; ctr$
   and  $\alpha_N \; \circ \; low$
   as the abstraction. 
  }
  \vspace{-3mm}
  \label{fig.TPFP.plot2}
\end{figure*}

The effect of using different abstractions and abstraction granularity (the $N$ parameter)
is shown in Figure \ref{fig.TPFP.plot2}. Based on the results in Figure \ref{fig.TPFP.plot1}, we now
use the lowest aggressiveness level (0). The graph of ${\sf uni}$ shows that increasing $N$
can greatly improve the FNN's ability to detect error, while keeping the false positive rate below
5\%. We also see $\alpha_N$ and $\alpha_N \; \circ \; low$ perform significantly better than
$\alpha_N \; \circ \; ctr$, implying that the choice of the abstraction function matters.
Compared to $\alpha_N$, $\alpha_N \; \circ \; low$ and $\alpha_N \; \circ \; ctr$ introduce
non-linear granularity. The results suggest that introducing more granularity in the region
(of $P$'s output) which are more error prone pays off.

%=====================================================================
%\section{Related Work} \label{sec.relatedwork}
%=====================================================================

%=====================================================================
\vspace{-8mm}
\section{Conclusion} \label{sec.concl}
%=====================================================================

\vspace{-3mm}
The experiment showed that, contrary to earlier attempts, it is 
possible to train Neural Networks, given an appropriate abstraction,
to become an artificial specification for a non-trivial program with acceptable precision.
As future work, more case studies are needed to see how this generalizes.

\vspace{-3mm}
\bibliographystyle{splncs04}
\bibliography{NN}

\appendix

\section{Results on Individual Mutations} \label{apx.individual.results}

The table below shows each of the mutation used in our experiment for simulating
errors. The mutations are the same as originally used in \cite{vanmali2002}.

{\footnotesize
\begin{center}
\begin{tabular}{|l|l|} \hline
  {\em line}  & {\em mutation}  \\ \hline
  2: $\begin{array}{cl}
          & {\sf Region}==5 \\
       || & {\sf Region}==6 \end{array}$ &
     \begin{tabular}{l}
       {\sf M1:} \lstinline$Region==5$, \\
       {\sf M2:} \lstinline$Region==5 && Region==6$ \\
       {\sf M3:} \lstinline$Region==4 || Region==5$, \\
       {\sf M4:} \lstinline$Region==3 || Region==4$ \\ 
     \end{tabular} \\ \hline
  3: \lstinline$Age<18$ &
     \begin{tabular}{l}
       {\sf M5:} \lstinline$Age>18$, \\ {\sf M6:} \lstinline$Age<25$ \\
     \end{tabular} \\ \hline
  5: \lstinline$Citizenship==0$ &
     \begin{tabular}{l}
       {\sf M7:} \lstinline$Citizenship==1$ \\
     \end{tabular} \\ \hline
  7: \lstinline$State==0$ &
     \begin{tabular}{l}
       {\sf M8:} \lstinline$State==1$ \\
     \end{tabular} \\ \hline
  8: $\begin{array}{cl} & {\sf Region}==3 \\ || & {\sf Region}==4\end{array}$ &
     \begin{tabular}{l}
       {\sf M9:} \lstinline$Region==3$, \\ 
       {\sf M10:} \lstinline$Region==3 && Region==4$ \\
       {\sf M11:} \lstinline$Region==2 || Region==3$\ \\
       {\sf M12:} \lstinline$Region==1 || Region==2$ \\
     \end{tabular} \\ \hline
  11: \lstinline$Marital==0$ &
     \begin{tabular}{l}
       {\sf M13:} \lstinline$Marital==1$ \\
     \end{tabular} \\ \hline
  12: \lstinline$Dependents>0$ &
     \begin{tabular}{l}
       {\sf M14:} \lstinline$Dependents==0$ \\
       {\sf M15:} \lstinline$Dependents<0$ \\
     \end{tabular} \\ \hline
 15: \lstinline$Sex==0$ &
     \begin{tabular}{l}
       {\sf M16:} \lstinline$Sex==1$ \\
     \end{tabular} \\ \hline
  20: \lstinline$Marital==0$ &
     \begin{tabular}{l}
       {\sf M17:} \lstinline$Marital==1$ \\
     \end{tabular} \\ \hline
  21: \lstinline$Dependents>2$ &
     \begin{tabular}{l}
       {\sf M18:} \lstinline$Dependents>=2$ \\
       {\sf M19:} \lstinline$Dependents<2$ \\ 
       {\sf M20:} \lstinline$Dependents<=2$ \\
     \end{tabular} \\ \hline
  24: \lstinline$Sex==0$ &
     \begin{tabular}{l}
       {\sf M21:} \lstinline$Sex==1$ \\
     \end{tabular} \\ \hline
\end{tabular}
\end{center}
}

\begin{figure}
  \begin{center}
  \begin{tabular}{c}
  \includegraphics[trim={0, 35mm, 0, 30mm},clip, scale=0.35]{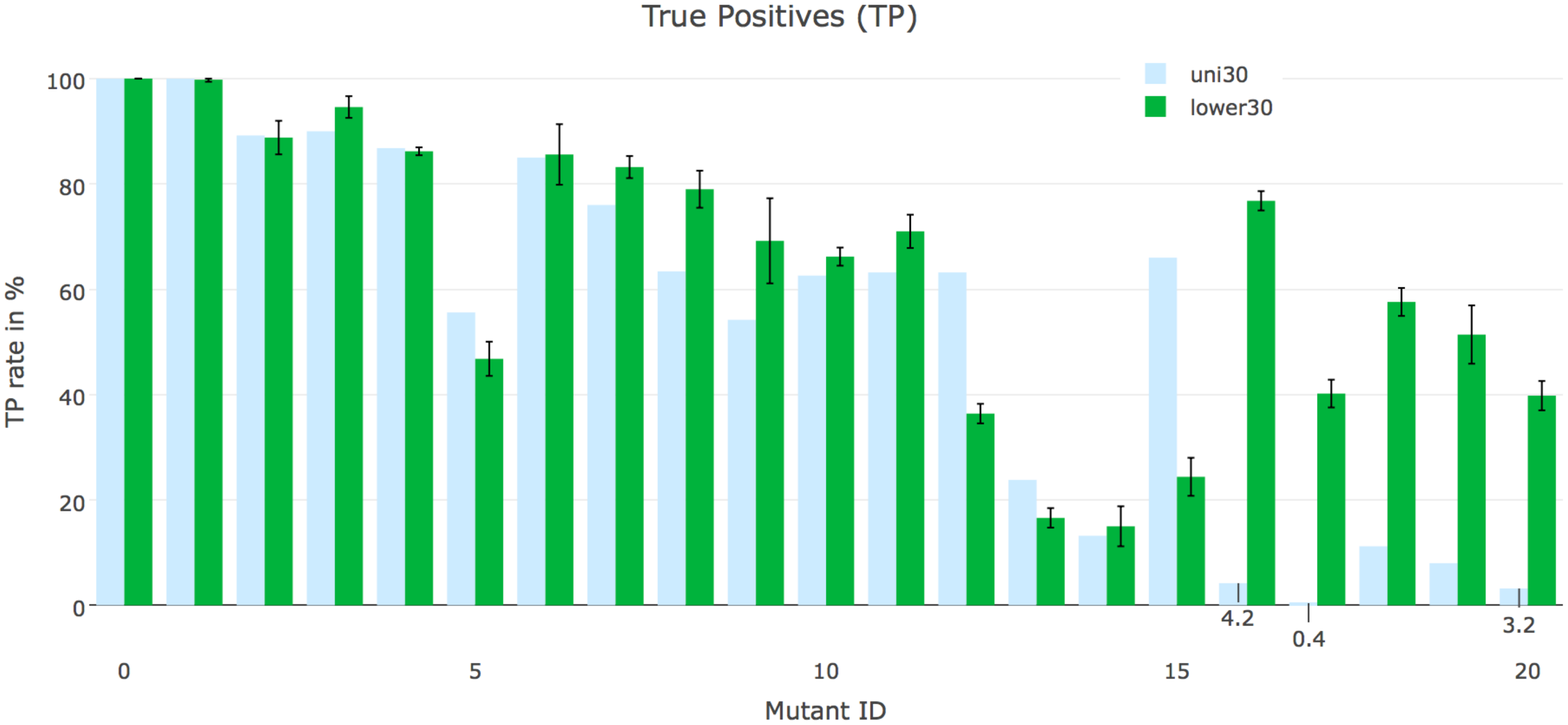} \\
  \includegraphics[trim={0, 35mm, 0, 30mm},clip, scale=0.35]{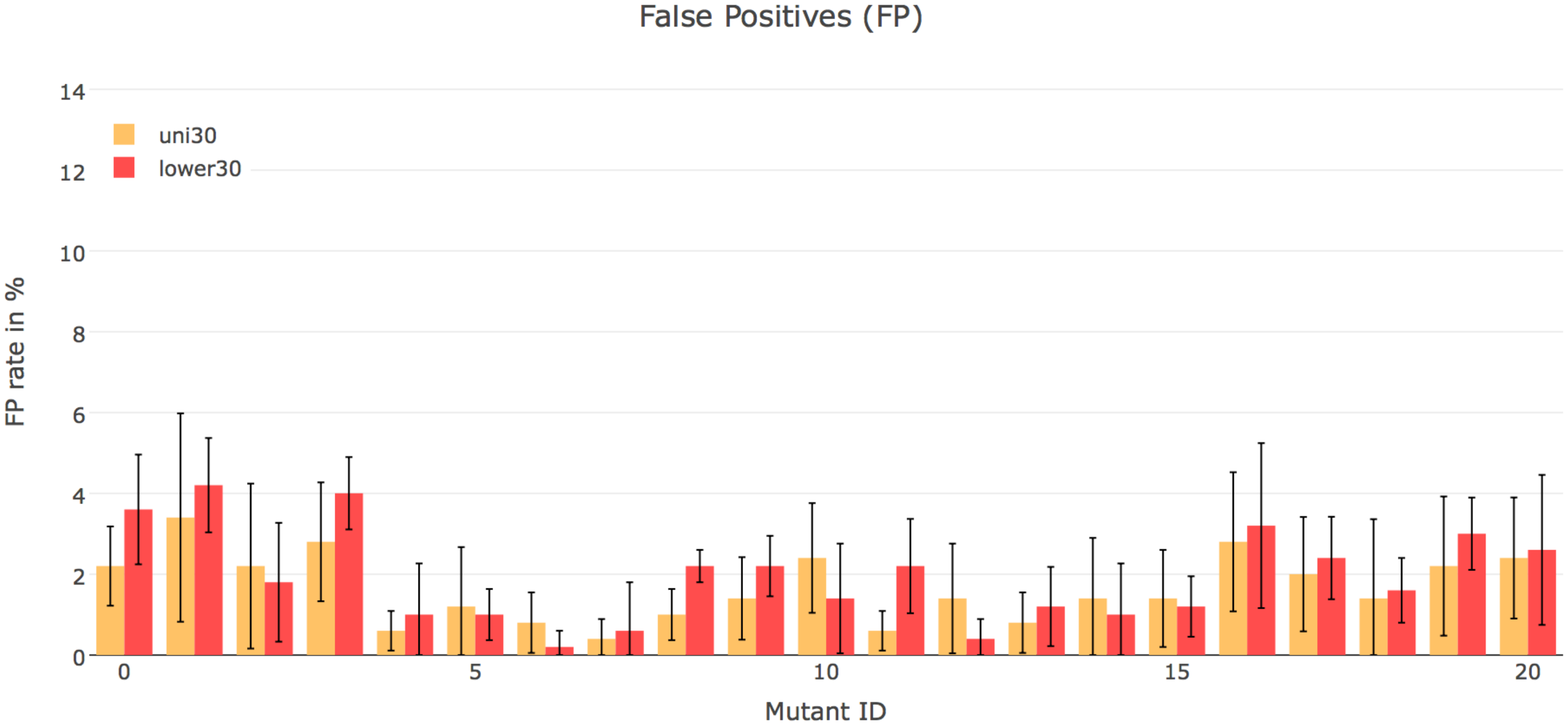}
  \end{tabular}
  \end{center}
  \vspace{-7mm}
  \caption{\em The graph at the top shows the individual true
    positive rate of the ${\sf uni}_{30}$ with its aggressiveness 
    level set to 0 (left bars) and that of the same FNN but using $\alpha_N  \circ low$
    as the abstraction function
    (right bars) on each mutant (M0..M20). The bottom graph shows the
    individual false positive rate.
    }
  \label{fig.TPFP.plot3}
\end{figure}

\noindent
Figure \ref{fig.TPFP.plot3} shows the true positive and false positive rates of the FNNs
on individual mutants. Two results of two FNNs are shown. The first is
${\sf uni}_{30}$ with its aggressiveness level set to 0; recall that
${\sf uni}_{30}$ uses the function $\alpha_N$ as the abstraction function. 
The second is ${\sf lower}_{30}$, with aggressiveness 0, but it uses $\alpha_N  \circ low$
as the abstraction.

Let's first consider the true positives (top graph).
We see here that on mutants
M16, M17, and M20 ${\sf uni}$ actually performs very poorly. See the
top graph in Figure \ref{fig.TPFP.plot3} ---$\sf uni$'s results on
these three cases are annotated in the graph.

In contrast, the hardest mutants for $\sf lower$ are M13 and M14,
but even on these mutants $\sf lower$ has a true positive rate of
${>}10\%$. Whether this 10\% is good enough depends on the situation.
We have defined the rate of true positives as the percentage of
wrong executions that the FNN judges as wrong as well. 
In particular, note that the metric is {\em not} defined as the percentage
of mutations that can be discovered. If we would define it like this,
$\sf lower$ would have 100\% rate of true positives because with enough
test cases eventually it will be able to detect all mutants.
So, $10\%$ individual rate of true positives for e.g. M14 means thus
that if we manage to trigger at least 10 distinct executions
that expose the mutation, statistically the FNN has a good chance to detect at least
one of them, and thus identifying the mutation. While this sounds very
encouraging, note that the actual probability for detecting the error also 
depends on the probability of producing executions that expose it. 
In the experiments, the probability of the latter is simply 1: we knew upfront
that there is a mutation, so generating the set of error exposing executions for each mutant
was not problematic. In a real regression testing setup, 
it is not possible to steer the testing process towards exposing a
particular error; we do not even know 
upfront if the new version of the program would contain any regression 
error at all. There are indeed tools to automatically generate test inputs
capable of generating a large number of test cases \cite{QuickCheck,prasetya2015t3}.
However, it is hard to generate test cases that are evenly distributed over all
control paths in the target program. Some paths may even be left uncovered 
because they are too difficult to cover, even by tools that employ more
sophisticated techniques like an evolutionary algorithm 
\cite{fraser2011evosuite} or symbolic calculation \cite{tillmann2008pex}.

The bottom graph in Figure \ref{fig.TPFP.plot3} shows the individual
false positive rate of $\sf uni$ and $\sf lower$. Each bar in
this graph also has its own error bar to indicate the standard
deviation $\sigma$ of the value the bar represents (the error bar is
capped above at 100 and below at 0, since true/false positive rates
can only range between [0..100]). For each mutant $M$, and each
experiment (e.g. $\sf lower$), the error $E_M$ of the false
positive rate of the experiment is calculated by randomly dividing
the set of 500 executions used in the experiment, into 5 bags of 100 elements and
then we calculate the false positive rate of the experiment with
respect to each bag. $E_M$ is defined as the standard deviation of
these values. The error bars indicate that sometimes the false
positive rate can peak above 5\%, though in average both
configurations, $\sf uni$ and $\sf lower$, produce rates that
are below $5\%$, for every individual mutant.

\end{document}